\documentclass[10pt,letterpaper]{article}

\usepackage{ccn}
\usepackage{graphicx}
\usepackage{amsmath}
\usepackage{amsfonts}
\usepackage{amssymb}
\usepackage{subcaption}
\usepackage{pslatex}
\usepackage{hyperref}
\usepackage{apacite} % 移除了 apacite，保留 natbib 以避免冲突
\usepackage{natbib}
\usepackage{doi}
\usepackage{flushend}

\title{Quantifying Entrainment Evidence: A Comparison of Frequentist and Bayesian Approaches for Information Processing Pathway Maps}

\author{
\begin{tabular}{cc} % 外层统一为一个两列的表格
\begin{tabular}{c}
{\large \bf Kaibo Zhang}\\
Department of Electronic Engineering\\
Tsinghua University\\
Beijing, China\\
\texttt{zkb21@mails.tsinghua.edu.cn}
\end{tabular}
&
\begin{tabular}{c}
{\large \bf Ji Wu}\\
Department of Electronic Engineering\\
Tsinghua University\\
Beijing, China\\
\texttt{wuji\_ee@tsinghua.edu.cn}
\end{tabular}
\\[4em]
\begin{tabular}{c}
{\large \bf Chao Zhang}\\
Department of Electronic Engineering\\
Tsinghua University\\
Beijing, China\\
\texttt{cz277@tsinghua.edu.cn}
\end{tabular}
&
\begin{tabular}{c}
{\large \bf Andrew Thwaites}\\
SHaPS\\
University College London\\
London, UK\\
\texttt{andrew.thwaites@ucl.ac.uk}
\end{tabular}
\end{tabular}
}

\begin{document}

\maketitle
\thispagestyle{plain} % 确保第一页有页码
\pagestyle{plain}     % 确保后续页面有页码

\begin{abstract}
Information Processing Pathway Maps (IPPMs) offer a scalable framework for formalizing the complex sequence of mathematical transformations applied to sensory stimuli. These maps chart the latency and cortical expression of computational steps, relying on statistical inference to link model outputs with observed neural activity. Traditionally, this mapping has relied on frequentist hypothesis testing. However, determining which of several competing computational models best explains neural data is a problem of model adjudication, arguably better suited to probabilistic inference. Here, we present a direct comparison between the established frequentist approach and a novel Bayesian framework for mapping cortical entrainment. While the Bayesian formulation retains the core strength of IPPMs—generating explicit predictions of time-varying neural signals—it fundamentally alters the selection criterion, shifting from rejecting a null hypothesis to quantifying the relative evidence for competing computational hypotheses. We evaluate the performance and interpretability of both approaches using an auditory neuroimaging dataset to reconstruct a known loudness-processing pathway. We discuss the implications of this shift for systems neuroscience, specifically regarding the handling of collinear models and the robust accumulation of evidence.
\end{abstract}

\section{Introduction}

A major ambition of cognitive neuroscience is to achieve a precise, mechanistic description of how the human brain processes information. Historically, researchers have faced a challenge in representing these complex processes in a way that is both mathematically rigorous and intuitively interpretable. Traditional approaches often fall short: precise mathematical representations can obscure the underlying neural concepts, while purely narrative descriptions lack the necessary formal detail.

\textit{Information Processing Pathway Maps} (IPPMs) bridge this gap by offering a clear, flexible, and mathematically accurate way to represent neural processing \citep{thwaites2025ippm}. They are formalised representations of the sequences of transformations that sensory information undergoes as it travels through the nervous system and cortex. Because IPPMs can be generated directly from neuroimaging data with high temporal resolution, such as electroencephalography (EEG) or magnetoencephalography (MEG), they serve as a scalable tool for reverse-engineering brain processes.

However, the validity of any resulting map rests entirely on the statistical criterion used to identify its nodes. To date, IPPM construction has relied on frequentist statistics, identifying processing stages by rejecting the null hypothesis that a model's output is uncorrelated with neural activity. While robust, this approach is conceptually indirect: it tells us the probability of the data assuming the model is irrelevant ($P(D|H_0)$), rather than the probability that the model is the correct description of the cortical process ($P(H|D)$). In this paper, we contest this standard. We introduce a Bayesian framework for IPPM inference and perform a side-by-side comparison with the traditional frequentist approach. We assess their respective abilities to recover a known pathway of auditory loudness processing, evaluating how each framework handles the critical task of arbitrating between competing computational descriptions of brain function.

While the Bayesian framework offers a probabilistic path for model adjudication, it is not without significant challenges that have long centered the frequentist-Bayesian debate. Critics often highlight the inherent subjectivity in the choice of priors, arguing that an ill-informed prior can lead to 'arbitrary' results that reflect the researcher's biases rather than the underlying data \citep{gelman1995bayesian, efron2013250}. Furthermore, Bayesian inference is highly sensitive to model misspecification; if the candidate model set does not contain a 'true' generator of the data, the posterior distribution can become misleadingly confident in the 'least-bad' option \citep{kass1995bayes}.

In contrast, the frequentist paradigm remains the standard for neuroimaging due to its robust control over False Positive Rates (Type I errors) without requiring prior assumptions. However, as the field shifts toward more complex representations of cortical processing, the limitations of binary null-hypothesis testing become more pronounced. This study does not advocate for the wholesale abandonment of frequentist rigor, but rather explores how a Bayesian approach—when grounded in physiologically-motivated priors—can provide a more nuanced quantification of evidence for competing neural transformations \citep{stephan2009bayesian, penny2004comparing}.

\subsection{Constructing Information Processing Pathway Maps}

The core assumption behind IPPMs is that the brain functions as an information processing system, transforming sensory input via a series of computational stages. The research goal is to reverse-engineer this system by testing which computational models best explain the observed neural activity across the cortex.

An IPPM is formally characterized as a directed acyclic graph (DAG), where:

\begin{itemize}
    \item \textbf{Nodes} represent a time-varying signal of some observation in the brain (e.g., source current) at a specific location.
    \item \textbf{Directed Edges} represent a computable mathematical transform that defines the relationship between two nodes (input to output).
    \item The \textbf{latency} of the nodes' signal being encoded (relative to the original sensory input) is denoted by their position on the x-axis.
\end{itemize}

\begin{figure*}
    \centering
    \includegraphics[width=1\linewidth]{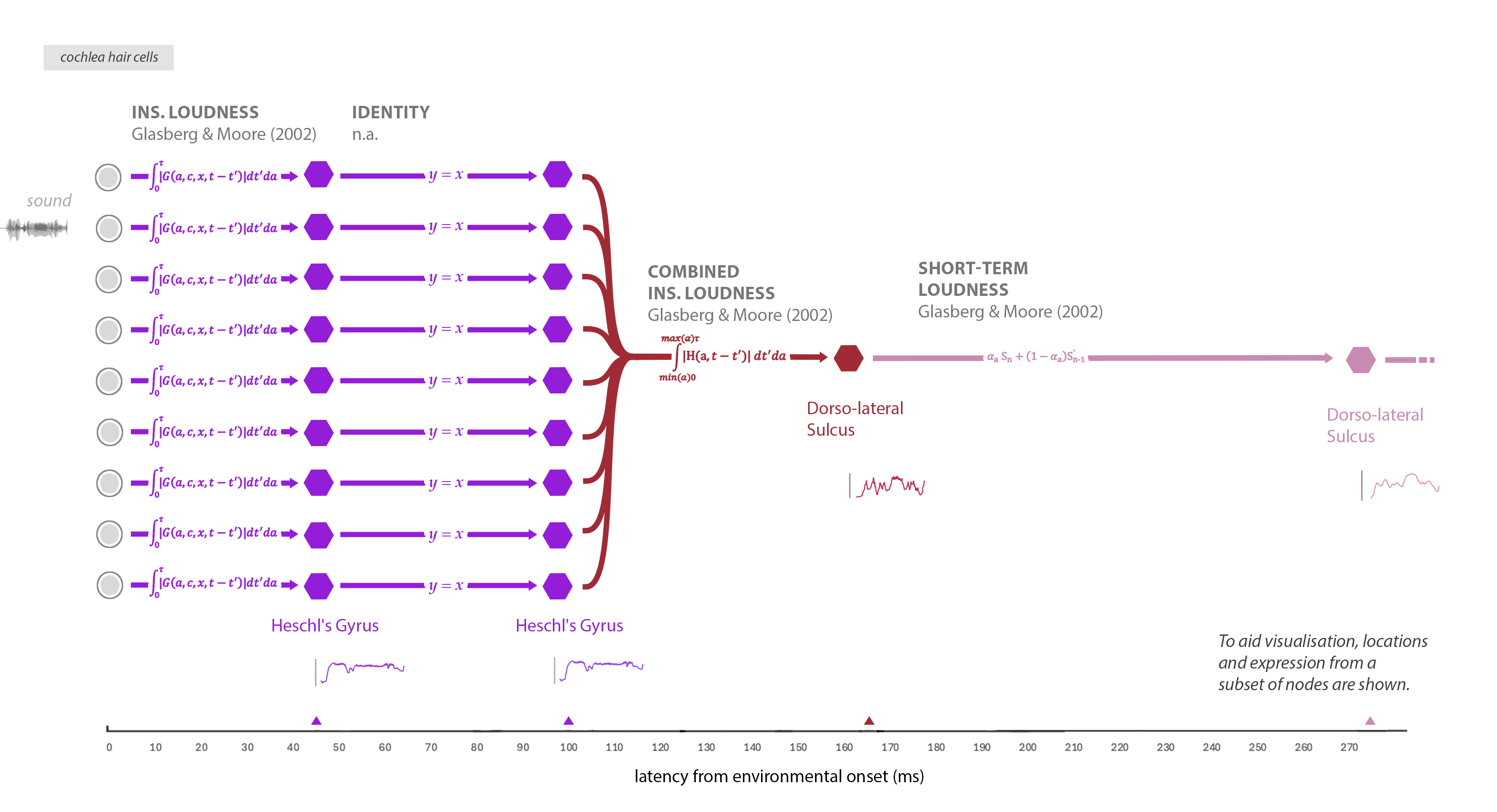}
    \caption{An (idealised) IPPM of loudness processing in the auditory cortex. Adapted from \citet{thwaites2017tonotopy}.}
    \label{fig:idealised-loudness-IPPM}
\end{figure*}

\noindent To create an IPPM, several prerequisites are required, following a general three-stage process:

\begin{enumerate}
    \item \textbf{Hypothesis Generation:} A set of plausible mathematical transformations ($H_1, H_2, \dots, H_n$)  are proposed, describing the conversion of a stimulus (or the output of a previous stage) into a specific time-varying signal in an encoding medium, $y=f(x)$. These transformations are typically drawn from well-established models, such as the Moore loudness model \citep{glasberg2002model} or the CIECAM02 model of color perception \citep{moroney2002ciecam02}.
    \item \textbf{Evidence Mapping (The Expression Plot):} For a given dynamic stimulus, each computational model component generates a specific time-series prediction (e.g., the instantaneous loudness contour over time). This prediction is then compared with the observed electrophysiological activity from various brain regions (hexels). The quality of fit (the evidence) between the predicted activity and the cortical activity is evaluated across a wide range of possible latencies. The results are visualized as an \textbf{expression plot}. This plot charts the evidence that any one region holds the output of a transform and the latency at which this occurs (see Figure \ref{fig:expression}).
    \item \textbf{Map Inference:} The IPPM itself is inferred from the temporal ordering and spatial locations identified in the expression plot (see Figure \ref{fig:idealised-loudness-IPPM}, cf.\, \citealp{lakra2025strategies}).
\end{enumerate}

\noindent A critical step in this methodology is Step 2: quantifying the measure of evidence for a hypothesized transform's output being expressed in a specific cortical region at a given latency. Existing IPPM studies have primarily relied on frequentist approaches, evaluating the goodness-of-fit using metrics like correlation and quantifying the evidence strength with statistically stringent methods, such as $\alpha$-thresholds adjusted via Šidák correction for multiple comparisons \citep{sidak1967rectangular}. Examples of this approach are found in studies of tonotopic loudness processing \citep{thwaites2017tonotopy}, color perception \citep{thwaites2018ciecam02}, and motion processing \citep{wingfield2025motion}.

In this paper, we address the question of the optimal evidence measure by comparing the robustness of these established frequentist approaches with a new formulation based on Bayesian inference. We provide a detailed methodology for basing the inference on Bayesian statistics, present a comparative IPPM, and discuss the implications of this alternative framework for the construction and interpretation of Information Processing Pathway Maps.

\section{The Traditional Frequentist Approach}

This measure of plausibility is typically quantified by a p-value, which is the probability of observing data as extreme, or more extreme, than the data actually observed, assuming the null hypothesis $H_0$ is true (i.e., that there is no true entrainment to the computational signal in that location in the brain). This is the classic frequentist approach.

The evidence for entrainment is typically quantified by comparing the distribution of the observed fit (e.g., correlation) with a suitable null distribution. The null hypothesis $H_0$ is rejected if the evidence supporting the match (the observed fit) surpasses a predefined significance threshold $\alpha$.

Despite its demonstrated robustness, the traditional frequentist approach suffers from a fundamental conceptual limitation. The metric for evidence, the $p$-value, quantifies the probability of the data given the null hypothesis $P(D \vert H_0)$, which is formally not what is of primary interest to the researcher. What we ultimately want is the probability of the hypothesis given the data $P(H \vert D)$---that is, the actual support for the computational hypothesis being true.

\section{Migrating to a Bayesian approach}

We propose casting IPPM construction as a formal Bayesian model comparison problem. Unlike the frequentist approach, which evaluates entrainment against a default null hypothesis of zero correlation, the Bayesian framework evaluates the relative evidence across a closed, explicitly defined hypothesis space, $H = \{H_0, H_1, ..., H_n\}$. Here, $H_1$ through $H_n$ represent the candidate computational transforms (e.g., intermediate auditory features like instantaneous loudness), and $H_0$ is the null model predicting a constant output. This approach builds upon the established use of Bayesian Model Selection (BMS) in neuroimaging, which allows for the direct adjudication between competing computational hypotheses without being restricted to nested models \citep{stephan2009bayesian, rosa2010bayesian}.

\subsection{Priors and Physiological Constraints}

Following standard Bayesian inference \citep{jaynes2003probability, knill2004bayesian}, each hypothesis is assigned a prior probability, $P(H_i)$. A distinct aspect of this framework for IPPM construction is the ability to use biologically informed priors—such as known cortical latencies or sensor modalities—to constrain the hypothesis space. However, to establish a baseline comparison against the frequentist method in this initial study, we employ a uniform prior across all models, isolating the effect of the likelihood function on model selection.

\subsection{Likelihood and Evidence Mapping}

The evidence for a specific computational model $H_i$ at a given cortical latency $l$ is quantified via the likelihood $P(D|H_i, l)$. This is computed assuming a Gaussian noise distribution over the residual sum of squares between the observed neural signal $D$ and the time-varying prediction generated by the model.

\subsection{Map Inference via Posterior Probability}

Rather than relying on a fixed $p$-value threshold, we evaluate the transformations by computing the posterior probability $P(H_i|D,l)$. Because the hypothesis space is treated as a closed system, the posterior effectively enforces a competition for probability mass among the candidate models (including the null). The resulting Bayesian expression plot maps the temporal dynamics of information processing by charting the posterior support across time, identifying an optimal processing node at the latency where a transform's posterior probability is maximized.

\section{Experimental validation}

\subsection{EMEG dataset} 

We compared both the Frequentist and Bayesian approaches to IPPMs on the same dataset, the Kymata-SOTO-English dataset (\cite{yang2026kymata}).  Both approaches used the same pre-processing pipeline - only the statistics framework changed. This dataset is comprised of 20 native-English-speaking individuals who are listening to an English-language podcast on ice-cream. This naturalistic stimulus contains momentary fluctuations in the sound intensity over the course of the podcast, allowing us to re-create a loudness IPPM (see next section).

\subsection{The Glasberg--Moore model of loudness}

To validate the Bayesian framework, we attempted to reconstruct the IPPM for loudness processing. This specific pathway was established experimentally in \citet{thwaites2017tonotopy} and is based on the perceptual model of loudness proposed by \citet{glasberg2002model}. The model hypothesizes that loudness perception arises from three sequential processing stages (see Figure \ref{fig:idealised-loudness-IPPM}).

In the first stage, simulating peripheral processing, sound entering the cochlea is decomposed into distinct frequency bands via an array of level-dependent bandpass filters. The output of each filter undergoes compressive nonlinearity, mimicking the mechanics of the basilar membrane, to generate a quantity known as specific loudness. For the purposes of cortical mapping, this stage is modeled using nine distinct frequency channels (referred to as `IL1--9'), each spanning a bandwidth of 4 Cams on the ERB$_N$-number scale. Neuroimaging evidence locates the expression of these channel-specific signals in Heschl's Gyrus at latencies of approximately 45 ms and 100 ms.

In the second stage, these frequency-specific channels are integrated---summed across the spectrum---to form a single, aggregate measure known as instantaneous loudness (`IL'). This spectral integration is observed in the dorso-lateral sulcus (DLS) at a latency of roughly 165 ms.

In the third stage, the instantaneous loudness undergoes temporal integration. The signal is smoothed over time using a running average, similar to an automatic gain control system, to produce short-term loudness (`STL'). This final perceptual construct is expressed at a latency of approximately 275 ms, located bilaterally in the DLS and the superior temporal sulcus.

The high replicability of this specific IPPM has been demonstrated in independent cohorts \citep{yang2026replication}, establishing it as a robust `ground truth' benchmark for methodological validation. Because the sequence of transforms and their expression latencies are well-characterized, we can confidently use this pathway to assess whether the Bayesian framework recovers known signal structures or introduces novel artifacts.

To mitigate computational demands, we applied this analysis to sensor-level data rather than source-reconstructed data. While this approach precludes precise cortical localization, it allows for the robust validation of the identified transforms and their associated latencies.

\subsection{Procedure}

We generated expression plots for the 11 transforms of the Glasberg-Moore model of loudness, using both the original frequentist method, and the new Bayesian method. The only difference between these two plots is how `good entrainment' is quantified: $p$-values for the former, probability for the latter.

From both of these, we also use the same automated procedure to infer an IPPM \citep{lakra2025strategies}.

\section{Results}

\subsection{Expression plots}

\begin{figure*}[ht!]
    \centering
    \begin{subfigure}[t]{0.4\textwidth}
        \centering
        \includegraphics[width=\textwidth]{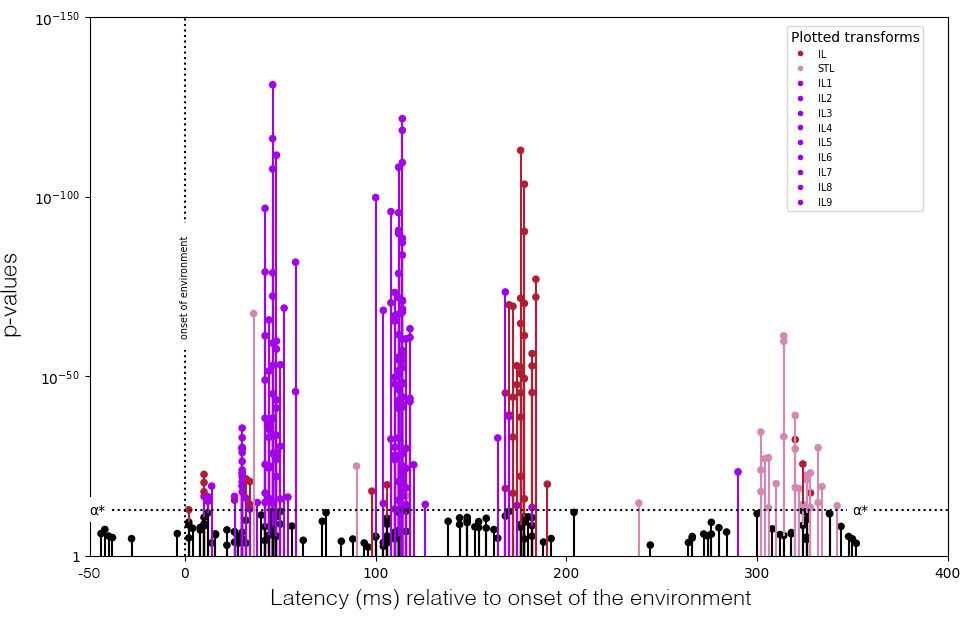}
        \caption{Frequentist expression plot}
        \label{fig:expression-frequentist}
    \end{subfigure}%
    ~
    \begin{subfigure}[t]{0.4\textwidth}
        \centering
        \includegraphics[width=\textwidth]{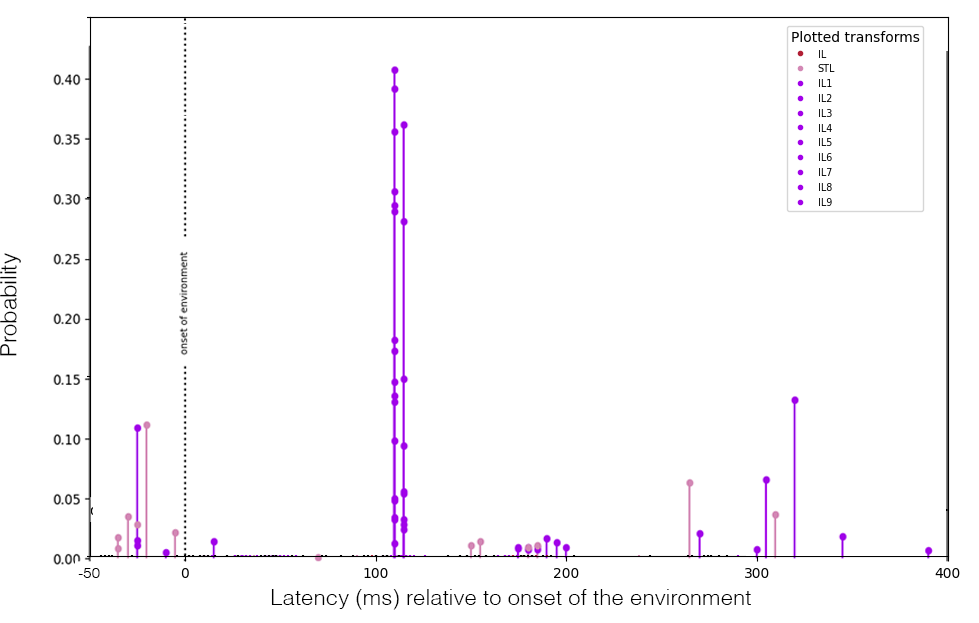}
        \caption{Bayesian expression plot}
        \label{fig:expression-bayesian}
    \end{subfigure}
    \caption{A comparison of the expression plots created with (a) frequentist and (b) Bayesian approaches. All 11 of the loudness transforms are plotted, coloured by stage, across processing lags from $-200$ to $+400$ ms, relative to the auditory environment. The evidence (either $p$-values in the frequentist case or probability in the Bayesian case) at a given channel, over all latencies, are marked as stems. In the frequentist case, stems at or crossing the p-threshold value ($p \approx 3 \times 10^{-13}$) indicate significant expression of the one of these functions at that latency.}
    \label{fig:expression}
\end{figure*}

Figure \ref{fig:expression} shows a comparison between the frequentist and Bayesian expression plots. We observed that both plots show similar ‘spikes’ of entrainment, especially at 100 and 300 ms latency. However, the Bayesian expression plot also appears to show some spikes before the 0 ms latency, and it is also missing the 50 ms and 180 ms latency spikes found in the frequentist version.

\subsection{Resultant IPPMs}

\begin{figure*}[ht!]
    \centering
    \begin{subfigure}[t]{0.4\textwidth}
        \centering
        \includegraphics[width=\textwidth]{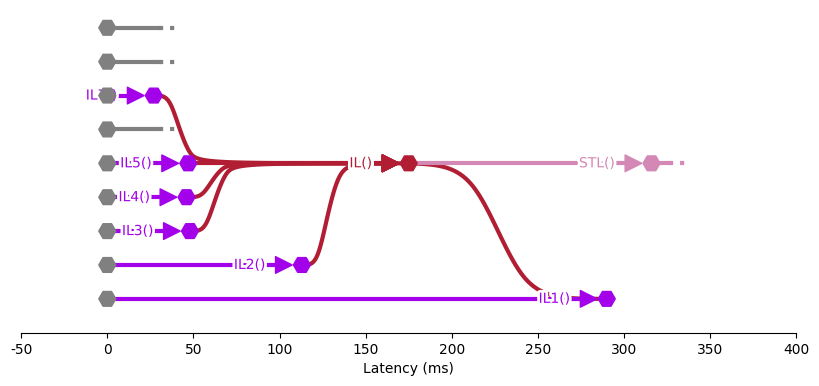}
        \caption{Frequentist IPPM}
        \label{fig:IPPM-frequentist}
    \end{subfigure}%
    ~
    \begin{subfigure}[t]{0.4\textwidth}
        \centering
        \includegraphics[width=\textwidth]{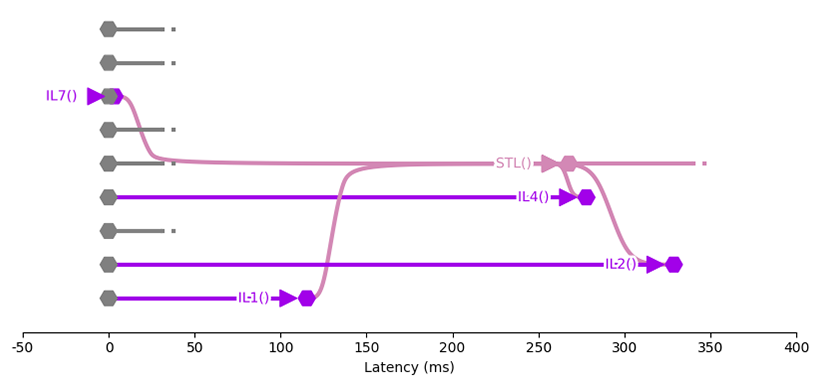}
        \caption{Bayesian IPPM}
        \label{fig:IPPM-bayesian}
    \end{subfigure}
    \caption{A comparison of the IPPMs inferred from the expression plots in Figs \ref{fig:expression-frequentist} and \ref{fig:expression-bayesian}. The same latency x-axis, and colouring scheme, are used as Fig \ref{fig:expression}. }
    \label{fig:ippm}
\end{figure*}

Following the generation of both expression plots, we then put both sets of expression data into the IPPM inference procedure. Figure \ref{fig:ippm} shows a comparison between the resulting frequentist and Bayesian IPPMs. As one would expect, the two IPPMs show some similarity (particularly for the short-term loudness and channel instantaneous loudnesses), but the Bayesian IPPM is missing the combined instantaneous loudness node at 180 ms (present in the frequentist version), and it also charts some entrainment of channel specific loudnesses before 0 ms.

\section{Discussion}

\subsubsection{The Correspondence Between Likelihood and Posterior Probability} The broad similarity between the frequentist and Bayesian expression plots is not unexpected. While the two frameworks differ philosophically, the mathematical quantities they rely on—the likelihood of the data given the hypothesis, $P(D|H)$, and the posterior probability of the hypothesis given the data, $P(H|D)$ —are frequently highly correlated. In a Bayesian framework, the posterior is proportional to the product of the likelihood and the prior ($P(H|D)\propto P(D|H)P(H)$). When the prior distribution $P(H)$ is uniform (flat)—as was the case in our analysis, where we assumed no initial preference for any loudness transform—the shape of the posterior distribution is determined entirely by the likelihood. Under these conditions, the transform that maximizes the likelihood (the "best fit" in a frequentist sense) will mathematically also be the transform with the maximum a posteriori (MAP) probability. Consequently, in scenarios where the signal-to-noise ratio is high and the winning model is distinct, frequentist significance testing (which relies on rejecting the null based on likelihood-derived statistics) and Bayesian inference will often converge on the same conclusion. This mathematical coupling explains why p-values have historically served as effective, albeit indirect, proxies for evidential support in systems neuroscience. Note that while the frequentist approach yields a sparse and physiologically expected map, the Bayesian results in this initial implementation exhibit anti-causal artifacts and missing components (e.g., at 180ms). These results highlight the sensitivity of Bayesian inference to factors such as signal scaling and the necessity of well-tuned priors. Rather than demonstrating immediate superiority, these findings serve as a pilot investigation into the technical requirements for Bayesian IPPMs, illustrating how naive implementations can be susceptible to noise in high-dimensional EMEG data.

\subsubsection{The Role of the Hypothesis Space} Despite this theoretical convergence, Figure \ref{fig:expression} demonstrates that the two approaches can yield distinct results in practice, with spikes of entrainment moving or disappearing in the Bayesian map. This divergence stems from the fundamental difference in how the two frameworks treat the hypothesis space. The frequentist approach effectively poses a series of independent binary questions: \textit{Is this specific transform significantly better than chance?} If multiple collinear transforms (e.g., neighboring frequency channels) all fit the data well, the frequentist approach may flag all of them as significant.

In contrast, the Bayesian framework treats the hypothesis space as a closed system where probability mass is a finite resource that must be allocated among the candidates (plus the null). This creates a competitive environment: if four different transforms generate similar predictions that all fit the data well, they do not reinforce each other; instead, they "dilute" the probability mass, potentially leaving each with a modest posterior probability (e.g., ~25\%). Conversely, a unique transform with a distinct prediction might garner a higher posterior probability even if its absolute fit is slightly lower, simply because it faces no competition from similar models. This "Occam's razor" effect is a deliberate feature of Bayesian inference, forcing researchers to be explicit about their full set of hypotheses \citep{jaynes2003probability}. As the number of redundant hypotheses increases, this dilution effect becomes more pronounced, altering the topology of the resulting IPPM.

\subsubsection{Evaluating Model Fidelity}

While the theoretical properties of the Bayesian framework are compelling, the practical results presented in Figure \ref{fig:ippm} compel us to ask which map more accurately represents the ground truth of cortical processing. The presence of entrainment spikes prior to 0 ms in the Bayesian IPPM is particularly telling. Since cortical processing cannot causally precede the sensory stimulus, these spikes represent clear false positives—instances where the model has overfitted to noise that happened to correlate with the predictor. Furthermore, the absence of the 180 ms combined-loudness node—a robust feature identified in previous literature —suggests a higher Type II error rate (false negatives) in our current Bayesian implementation.

However, these shortcomings should not be viewed as an indictment of Bayesian inference itself, but rather as a reflection of the specific priors and likelihood functions employed in this initial comparison. The frequentist approach implicitly filters out pre-stimulus noise by treating each time-point independently and relying on a strict significance threshold. The Bayesian model, as implemented here with uniform priors, lacks this "common sense" constraint. A key strength of the Bayesian approach is precisely that we can formally encode constraints to mitigate such errors. By modifying the prior P(H) to account for factors such as sensor modality, cortical source depth, or well-established physiological latency constraints, we can significantly improve the map's validity. Thus, the noise susceptibility seen here is not a failure of the framework, but a demonstration of the necessity for biologically informed priors that go beyond the uniform assumptions used in this baseline comparison.

\subsubsection{The Flexibility of the Bayesian Framework}

Beyond the immediate comparison of results, the Bayesian framework offers a theoretical elegance and flexibility that the frequentist approach struggles to match. The traditional approach relies on a fixed decision rule (the p-value threshold) and a specific null hypothesis (zero correlation). In contrast, the Bayesian framework allows researchers to explicitly model the generative process of the signal.

For instance, where standard regression assumes homoscedastic Gaussian noise, a Bayesian formulation can easily be extended to model specific noise structures known to plague M/EEG data, such as alpha-band oscillations or cardiac artifacts. Furthermore, as the field of Information Processing Pathway Maps matures, we need not rely on "flat" priors. We can incorporate prior knowledge from independent studies—using the posterior latencies from one experiment as the prior distributions for the next. This capability transforms IPPM construction from a series of isolated experiments into a cumulative science, where evidence is systematically updated rather than restarted from scratch. This ability to quantify the relative probability of competing models ($(P(H_A|D)$ vs. $P(H_B|D)$) provides a far more intuitive metric for model selection than comparing p-values, which are not valid measures of effect size or model probability.

\subsection{Methodological Limitations}

A significant limitation of the current work is the absence of controlled simulations using synthetic datasets. In applied statistics, such simulations are essential for establishing a 'ground truth' and defining the precise failure modes of a new methodology. In this study, our 'ground truth'—the loudness processing pathway—is derived from prior IPPM literature \citep{thwaites2015tracking}, which itself utilized frequentist paradigms. This introduces a risk of circularity in our model adjudication.

We view the current work as a feasibility study, demonstrating how Bayesian inference handles the high-dimensional noise and complex temporal structures inherent in real EMEG recordings. Future research should prioritize the development of generative models capable of producing synthetic neural entrainment data. Such a framework would allow for a rigorous 'Experiment A' to quantify the sensitivity and specificity of the Bayesian posterior against an unquestionable ground truth, particularly in cases of model misspecification.

\subsubsection{Complexity and the Marginal Likelihood}

The current framework utilizes relatively simple parameterizations to establish a baseline comparison against frequentist methods. However, the most principled approach to Bayesian model selection involves computing the marginal likelihood (model evidence) by integrating the posterior over all nuisance parameters. During our investigation, we explored two such marginalization strategies: (1) integrating analytically over the noise variance, and (2) marginalizing over the signal amplitude scaling parameter.

While mathematically elegant, applying these fully marginalized likelihoods to high-dimensional EMEG data revealed significant challenges. Because EMEG data contains a massive number of time samples, the marginal likelihood becomes hypersensitive. In our tests, this hypersensitivity caused the posterior probabilities to collapse into extreme values (0 or 1), heavily penalizing any slight mismatch in signal shape and amplifying minor correlations with background noise, resulting in a highly erratic expression map. This indicates that to successfully leverage the full marginal likelihood without yielding extreme entries, the framework will require highly structured, hierarchical spatio-temporal priors to regularize the model space. We highlight the development of such bespoke, regularizing noise models as a critical direction for future Bayesian IPPM research.

\subsection{Enhanced Model Explorations}

To explore the potential for enhancing the IPPM framework, we conducted pilot investigations into two more targeted Bayesian modeling techniques: the use of literature-informed temporal priors and marginalized likelihoods.

First, we implemented a set of Gaussian priors centered on the established physiological latencies of the auditory pathway (e.g., 75 ms for early cortical processing, 165 ms for spectral integration, and 275 ms for temporal integration). Second, we explored a marginalized likelihood approach to treat signal amplitude and noise variance as nuisance parameters.

Our results from these investigations revealed that while these methods are theoretically principled, they introduce a high degree of sensitivity when applied to high-dimensional EMEG time series. Without additional hierarchical regularization, the marginalized posterior often exhibits 'probability collapse' or erratic expression patterns due to the model’s extreme sensitivity to minor noise correlations. Consequently, while we demonstrate the feasibility of these advanced structures, this study maintains a baseline model with uniform priors to ensure a direct and stable comparison with the established frequentist approach. These pilot findings underscore the necessity of developing bespoke, regularizing noise models as a prerequisite for fully realized Bayesian IPPMs.

\subsection{Generalizability and Multi-Dataset Application}

While this study serves as a proof-of-concept using a specific auditory EMEG dataset \citep{thwaites2015tracking}, the Bayesian IPPM framework is designed for broad applicability across different sensory modalities and recording techniques. Future validation should extend to diverse open-source repositories to test the framework’s robustness. For instance, the Kymata Atlas datasets \citep{yang2026kymata} provide high-quality speech-processing recordings that could test how Bayesian model selection adjudicates between complex linguistic transforms. Similarly, the CPEP (Cortical Processing of Evoked Potentials) dataset or the HCP (Human Connectome Project) Meg-Language recordings would allow for the evaluation of the framework in the context of visual and cross-modal integration. Extending the analysis to these varied data structures will be essential to determine if the Bayesian sensitivity observed here is a constant feature of the method or specific to the signal-to-noise ratios of auditory entrainment.

\subsection{Conclusion and future directions}

The endeavour to map computational models of sensory processing onto human cortical activity lies at the heart of cognitive computational neuroscience. Information Processing Pathway Maps (IPPMs) provide a scalable framework for this task, allowing researchers to formalize the sequence of mathematical transformations that underpin perception. However, as this framework expands to cover increasingly complex cognitive domains, the statistical engines driving these maps must be subjected to rigorous scrutiny.

In this work, we presented a direct comparison between the established frequentist approach and a novel Bayesian formulation for inferring these pathways. Our results highlight that the choice of statistical framework is not merely a technical implementation detail, but a fundamental decision about how we define and evaluate evidence in systems neuroscience. While frequentist frameworks are fully capable of modeling sophisticated noise structures (e.g., via generalized least squares or mixed-effects models), a Bayesian formulation offers a highly cohesive approach to handling the specific noise profiles known to plague M/EEG data, such as alpha-band oscillations or cardiac artifacts. Rather than relying on point estimates for these complex noise structures, a fully realized Bayesian framework allows researchers to treat them as nuisance parameters. By defining explicit priors for these physiological artifacts, researchers can seamlessly integrate and marginalize them out of the posterior distribution, naturally propagating our uncertainty about the noise into the final model evaluation.

We showed that while the Bayesian approach currently faces challenges regarding noise sensitivity—manifesting here as physiologically implausible pre-stimulus entrainment—it offers a unique pathway to resolving the ambiguity of the hypothesis space. By explicitly quantifying the relative probability of competing models ($P(H|D)$), rather than relying on the rejection of a null hypothesis (P($D|H_0$)), the Bayesian framework aligns more closely with the researcher's ultimate goal: determining which computational algorithm is most likely being implemented by the cortex.

This study investigates the feasibility and current challenges of implementing a Bayesian framework for IPPMs. We explore the trade-offs between frequentist robustness and Bayesian sensitivity, identifying key implementation hurdles in the transition to probabilistic model selection. It moves us away from a binary paradigm of significance testing and toward a cumulative science of cortical mapping, where specific noise assumptions can be marginalized out and where prior knowledge from the literature can be formally integrated. As we refine these methods, the ability to treat today’s posterior maps as tomorrow’s priors promises to transform IPPMs from isolated experimental snapshots into a unified, continuously updating atlas of human information processing.

\section{Acknowledgements}

This work was supported by a grant from Neuroverse FZCO to AT and CZ and a National Natural Science Foundation of China (NSFC) grant to CZ (62476151).

\newpage % 强行在这里切栏，让后面的内容从右侧那一栏（新 Column）顶端开始

\bibliographystyle{ccn_style}

\bibliography{bib}

@article{glasberg2002model,
  title={A model of loudness applicable to time-varying sounds},
  author={Glasberg, Brian R. and Moore, Brian C. J.},
  journal={Journal of the Audio Engineering Society},
  volume={50},
  number={5},
  pages={331--342},
  year={2002},
  publisher={Audio Engineering Society}
}

@book{jaynes2003probability,
  title={Probability Theory: The Logic of Science},
  author={Jaynes, Edwin T.},
  year={2003},
  publisher={Cambridge University Press}
}

@article{lakra2025strategies,
    author = {Lakra, Anirudh and Wingfield, Cai and Zhang, Chao and Thwaites, Andrew},
    title = {Strategies for automatic generation of information processing pathway maps},
    journal = {Frontiers in Neuroimaging},
    year = {2025},
    volume = {4},
    number = {1608390},
    doi = {10.3389/fnimg.2025.1608390}
}

@inproceedings{moroney2002ciecam02,
  title={The {CIECAM02} color appearance model},
  author={Moroney, Nathan and Fairchild, Mark and Hunt, Robert and Li, Changjun},
  year={2002},
  booktitle={{10th Color and Imaging Conference Final Program and Proceedings}},
  page={23--27},
  publisher={Society for Imaging Science and Technology}
}

@article{sidak1967rectangular,
    author = {Zbyněk Šidák},
    title = {Rectangular Confidence Regions for the Means of Multivariate Normal Distributions},
    journal = {Journal of the American Statistical Association},
    volume = {62},
    number = {318},
    pages = {626--633},
    year = {1967},
    doi = {10.1080/01621459.1967.10482935},
}

@article{thwaites2015tracking,
    author = {Thwaites, Andrew and Nimmo-Smith, Ian and Fonteneau, Elisabeth and Patterson, Roy D. and Buttery, Paula and Marslen-Wilson, William D.},
    title = {Tracking cortical entrainment in neural activity: auditory processes in human temporal cortex},
    journal = {Frontiers in Computational Neuroscience},
    year = {2015},
    volume = {9},
    number = {5},
    doi = {10.3389/fncom.2015.00005}
}

@article{thwaites2018ciecam02,
title = {Entrainment to the {CIECAM02} and {CIELAB} colour appearance models in the human cortex},
journal = {Vision Research},
volume = {145},
pages = {1--10},
year = {2018},
issn = {0042-6989},
doi = {10.1016/j.visres.2018.01.011},
author = {Andrew Thwaites and Cai Wingfield and Eric Wieser and Andrew Soltan and William D. Marslen-Wilson and Ian Nimmo-Smith}
}

@article{thwaites2025ippm,
    author = {Thwaites, Andrew and Zhang, Chao and Woolgar, Alexandra},
    title = {Information processing pathway maps — A scalable framework for mapping cortical processing},
    journal = {{NeuroImage}},
    year = {2025},
    volume = {317},
    number = {15},
    doi = {10.1016/j.neuroimage.2025.121345}
}

@article{wingfield2025motion,
  title={Tracking cortical entrainment to stages of optic-flow processing},
  author={Wingfield, Cai and Soltan, Andrew and Nimmo-Smith, Ian and Marslen-Wilson, William D. and Thwaites, Andrew},
  journal={Vision Research},
  volume={226},
  pages={108523},
  year={2025},
  doi={10.1016/j.visres.2024.108523}
}

@article{thwaites2017tonotopy,
    title = {Tonotopic representation of loudness in the human cortex},
    journal = {Hearing Research},
    volume = {344},
    pages = {244-254},
    year = {2017},
    issn = {0378-5955},
    doi = {10.1016/j.heares.2016.11.015},
    author = {Andrew Thwaites and Josef Schlittenlacher and Ian Nimmo-Smith and William D. Marslen-Wilson and Brian C.J. Moore},
}

@article{yang2026replication,
    author = {ChenTianyi Yang and Cai Wingfield and Jinyu Dong and Sharon Yuen Shan Ho and Anirudh Lakra and Alexandra Woolgar, Chao Zhang and Andrew Thwaites},
    year = {2026},
    title = {Replicability of Auditory Information Processing Pathway Maps Over Multi-Site Magnetoencephalography and ECoG Recordings},
    journal = {Preprint},
    url = {https://kymata.org/assets/preprints/replicating_TVL_across_datasets.pdf}
}

@article{yang2026kymata,
  title={Kymata Soto Language Dataset: an electro-magnetoencephalographic dataset for natural speech processing},
  author={Yang, ChenTianyi and Parish, Oliver and Klimovich-Gray, Anastasia and Wingfield, Cai and Marslen-Wilson, William D and Zhang, Chao and Woolgar, Alexandra and Thwaites, Andrew},
  journal={Scientific Data},
  year={2026},
  doi = {10.1038/s41597-026-06579-8},
  publisher={Nature Publishing Group UK London}
}

@article{knill2004bayesian,
  title={The Bayesian brain: the role of uncertainty in neural coding and computation},
  author={Knill, David C and Pouget, Alexandre},
  journal={TRENDS in Neurosciences},
  volume={27},
  number={12},
  pages={712--719},
  year={2004},
  publisher={Elsevier}
}

@article{rosa2010bayesian,
  title={Bayesian model selection maps for group studies},
  author={Rosa, MJ and Bestmann, Sven and Harrison, L and Penny, W},
  journal={Neuroimage},
  volume={49},
  number={1},
  pages={217--224},
  year={2010},
  publisher={Elsevier}
}

@article{stephan2009bayesian,
  title={Bayesian model selection for group studies},
  author={Stephan, Klaas Enno and Penny, Will D and Daunizeau, Jean and Moran, Rosalyn J and Friston, Karl J},
  journal={Neuroimage},
  volume={46},
  number={4},
  pages={1004--1017},
  year={2009},
  publisher={Elsevier}
}

@book{gelman1995bayesian,
  title={Bayesian data analysis},
  author={Gelman, Andrew and Carlin, John B and Stern, Hal S and Rubin, Donald B},
  year={1995},
  publisher={Chapman and Hall/CRC}
}

@article{efron2013250,
  title={A 250-year argument: Belief, behavior, and the bootstrap},
  author={Efron, Bradley},
  journal={Bulletin of the American Mathematical Society},
  volume={50},
  number={1},
  pages={129--146},
  year={2013}
}

@article{kass1995bayes,
  title={Bayes factors},
  author={Kass, Robert E and Raftery, Adrian E},
  journal={Journal of the American Statistical Association},
  volume={90},
  number={430},
  pages={773--795},
  year={1995},
  publisher={Taylor \& Francis}
}

@article{penny2004comparing,
  title={Comparing dynamic causal models},
  author={Penny, William D and Stephan, Klaas E and Mechelli, Andrea and Friston, Karl J},
  journal={Neuroimage},
  volume={22},
  number={3},
  pages={1157--1172},
  year={2004},
  publisher={Elsevier}
}

\end{document}